\documentclass[aps,twocolumn, showpacs,prb,superscriptaddress,floatfix,shortbibliography,citeautoscript]{revtex4-1}
\usepackage{graphicx}
\usepackage[intlimits,tbtags]{amsmath}
\usepackage[dvipsnames]{xcolor}
\usepackage{amssymb}
\usepackage{array}
\usepackage{comment}
\usepackage{dcolumn}
\usepackage{bm}
\usepackage{multirow}
\usepackage{siunitx}
\usepackage[version=4]{mhchem}
\usepackage{tabularx}
\usepackage[colorlinks=true,linkcolor=blue]{hyperref}
\usepackage{cleveref}
\usepackage[export]{adjustbox}
\usepackage[normalem]{ulem}

\newcommand{\halle}{Institut f\"ur Physik, Martin-Luther-Universit\"at
  Halle-Wittenberg, D-06099 Halle, Germany}
  
\newcommand{\virginia}{Department of Physics, West Virginia University, Morgantown, WV 26506, USA}

\newcommand{\lux}{Department of Physics and Materials Science, University of Luxembourg, L-1511 Luxembourg, Luxembourg}
\newcommand{\bochum}{Research Center Future Energy Materials and Systems of the University Alliance Ruhr, Faculty of Mechanical Engineering, Ruhr University Bochum, Universitätsstraße 150, D-44801 Bochum, Germany}

\begin{document}

\author{Hai-Chen Wang}
\affiliation{\halle} 
\author{Ahmad W. Huran}
\affiliation{\halle} 
\author{Miguel A. L. Marques} 
\email{miguel.marques@rub.de}
\affiliation{\bochum}
\author{Muralidhar Nalabothula}
\affiliation{\lux}
\author{Ludger Wirtz}
\affiliation{\lux}
\author{Zachary Romestan}
\affiliation{\virginia}
\author{Aldo H. Romero}
\email{Aldo.Romero@mail.wvu.edu}
\affiliation{\virginia}
\affiliation{\lux}

\title{Two-Dimensional Noble Metal Chalcogenides in the Frustrated Snub-Square Lattice} 

\begin{abstract}


We study two-dimensional noble metal chalcogenides, with composition \ce{\{Cu, Ag, Au\}2\{S, Se, Te\}}, crystallizing in a snub-square lattice. This is a semi-regular two-dimensional tesselation formed by triangles and squares that exhibits geometrical frustration. We use for comparison a square lattice, from which the snub-square tiling can be derived by a simple rotation of the squares. The mono-layer snub-square chalcogenides are very close to thermodynamic stability, with the most stable system (\ce{Ag2Se}) a mere 7~meV/atom above the convex hull of stability. All compounds studied in the square and snub-square lattice are semiconductors, with band gaps ranging from 0.1 to more than 2.5~eV. Excitonic effects are strong, with an exciton binding energy of around 0.3~eV. We propose the Cu (001) surface as a possible substrate to synthesize {\ce{Cu2Se}}, although many other metal and semiconducting surfaces can be found with very good lattice matching.

\end{abstract}

\maketitle

\section{Introduction}
In the two-dimensional (2D) world, the plane has eleven different Euclidean tesselations using convex regular polygons. These symmetrical motifs have fascinated mankind for centuries and have been used as decorative elements since Roman and Islamic times or, more recently, in the beautiful work of M. C. Escher. Of these eleven, three are regular and are characterized by the number of edges meeting at each vertex, which can be either six (in the triangular lattice), three (in the hexagonal lattice), or four (in the square lattice). Some of the most notable materials in the atomic 2D world, such as graphene, the transition metal dichalcogenides, black phosphorus, etc., belong to this family. The remaining 8 lattices are semi-regular and are constructed from more than one regular polygon.

The trihexagonal tiling is perhaps the most studied semi-regular tesselation, often called the Kagome lattice, due to its use in traditional Japanese basketry. This motif can be found in the layers of some naturally occurring minerals, and the presence of the equilateral triangles leads to a geometrical frustration responsible for an exotic behavior of the electronic and magnetic properties. For example, 
kagome compounds, such as \ce{Fe3Sn2}~\cite{ye2018massive,Lin2018}, \ce{FeSn}~\cite{Kang2019}, \ce{YMn6Sn6}~\cite{Li2021}, or \ce{CoSn}~\cite{Huang2022} can exhibit Dirac cones and flat bands. Recently, a kagome material, \ce{KV3Sb5}, was found to have an unconventional chiral charge order, with a topological band structure and a superconducting ground state~\cite{Jiang2021}.

Here we are concerned by another, much less studied, semi-regular lattice, specifically the snub-square tiling. This tesselation consists of regular squares and triangles of matching edges, arranged so that exactly five edges meet at every vertex, and no edge is shared among two squares. Examples of this lattice can be found at larger length scales in two-dimensional metal–organic networks. For example, in Ref.~\onlinecite{cija2013} the snub-square tiling could be fabricated by performing the cerium-directed assembly of linear polyphenyl molecular linkers with terminal carbonitrile groups on an Ag(111) surface and by tuning the concentration and the stoichiometric ratio of rare-earth metal centers to ligands. This tesselation is also created by connecting a neutral rod-shaped secondary building unit with a cationic dicarboxylate ligand~\cite{Song2015} or by 
the linking of trans-{LnI$_2$}$^+$ nodes (Ln = Gd, Dy) by both closed-shell and anion radicals of 4,4'-bipyridine~\cite{Chen2021}.  Furthermore, in this latter case, the occurrence of sizable magnetic exchange interactions and slow relaxation of magnetization behavior was observed~\cite{Chen2021}. We emphasize that triangles in the snub-square lattice lead to a geometrical frustration of the lattice (as in the Kagome lattice), so we can expect unique electronic and magnetic properties.

The formation of these systems has also been investigated by computer simulations. Antlanger \textit{et al.} succeeded, using a bottom-up strategy, to decorate patchy particles so that they self-assemble in most Archimedean tilings~\cite{Antlanger2011}. Furthermore, they found that the snub square was stable at intermediate or even elevated pressure values due to its compact structure, involving only triangles and squares as building polygonal units. Reference~\onlinecite{Whitelam2016} has shown that the self-assembly of  Archimedean networks requires a combination of the geometry of the particles and chemical selectivity. Finally, the Archimedean tiling can be formed in mixtures of a pentavalent molecule and a linear linker, the driving force being the mobility of the linker~\cite{Baran2020}.

\begin{figure}[htb]
    \begin{center}
    \begin{tabular}{c}
    (a)~$P4/nmm$  \\[0.25cm]
    \includegraphics[width=5cm]{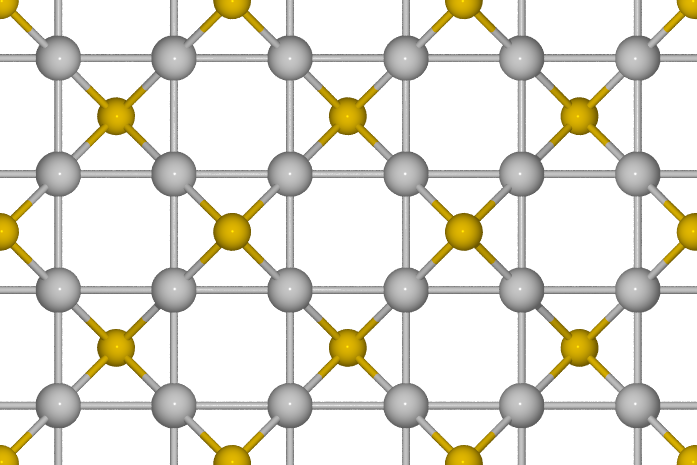} \\
    \includegraphics[width=5cm]{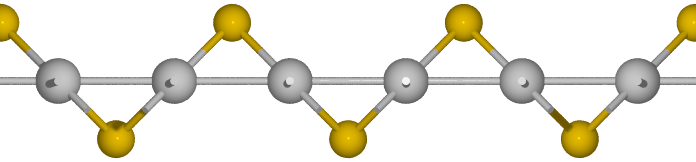} \\
    (b)~$P42_{1}2$ \\[0.25cm]
    \includegraphics[width=5cm]{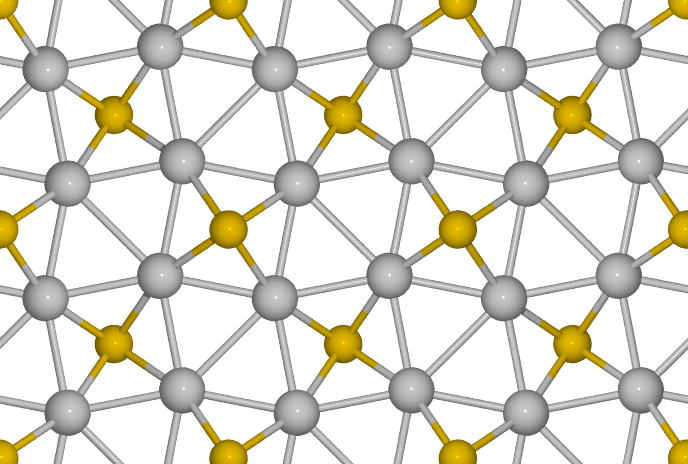} \\
    \includegraphics[width=5cm]{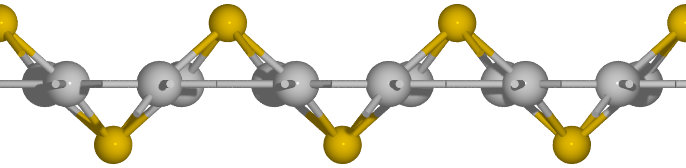}
    \end{tabular}
    \end{center}
    \caption{Crystal structures of 2D--\ce{Ag2Se} with top view and side view. The silver and yellow spheres denote Ag and Se atoms, respectively. The metallic framework forms either (a)~a square ($P4/nmm$) or a (b)~a snub-square lattice ($P42_{1}2$), with the chalcogen atoms alternating above and below the squares.}
    \label{fig:structure}
\end{figure}

Recently, it was discovered that the snub-square Archimedean lattice (as well as the undistorted square lattice form) can also exist in some group-IB-chalcogenides~\cite{Gao2021} with composition M$_2$Ch (where M is Cu, Ag, or Au and Ch is a chalcogen). The crystalline structure of these 2D systems is illustrated in Fig.~\ref{fig:structure}. The metal atoms are arranged in a flat snub-square lattice, while the chalcogen atoms are located at the center of the squares alternating above and below the plane. These systems are very close to thermodynamic stability, only slightly higher in energy than their bulk crystal phases. We also note that other 2D structures of Group-IB chalcogenides with similar stability have been predicted in Ref.~\onlinecite{Gao2021}. {The focus of our current study is the systematic comparison of the snub-square phases with the undistorted square-lattice phases.}

Related snub-square lattices were recently proposed for \ce{BaO3} and \ce{TiO2}~\cite{huran2021two}. In the former system, the Ba atoms form a planar snub-square lattice with \ce{O2} units filling both the triangular tiles (perpendicular to the plane) and the square tiles (in the plane). In the latter, the Ti sublattice is highly buckled, with the triangles decorated by one O atom and the squares decorated by two O atoms. However, in contrast with these two systems, 2D metal chalcogenides do not involve unusual metallic oxidation numbers (like in \ce{BaO3}) and are very close to the convex hull of thermodynamic stability (for comparison, the \ce{TiO2} system is 138~meV from the hull). As such, one can expect they should be much simpler to synthesize.

Here we will discuss in detail a series of group IB snub-square chalcogenide properties. {Specifically, we investigate the underlying bonding mechanism that in some cases stabilizes the snub-square phase as compared to the square one. We compare the electronic properties, optical absorption, vibrational properties, and Raman spectra between the two phases.} We also propose suitable substrates with minimal mismatch to grow 2D snub-square chalcogenides. Finally, we consider the possibility of making quasi-crystalline lattices based on these systems.

\section{Methods}
\label{sec:method}

The density-functional theory (DFT) calculations of optimized geometries and electronic band structures are performed via the Vienna \emph{ab initio} simulation package \textsc{VASP}~\cite{vasp1,vasp2} with the projector augmented wave method (PAW)~\cite{paw}. The plane-wave cutoff is set to 520~eV. A vacuum region of at least 15~\AA{ }is applied to the 2D slabs and the geometries are optimized until the forces are smaller than 0.005~eV/\AA.

The Brillouin zones are sampled by Monkhorst-Pack $k$-grids centered at $\Gamma$, while the densitiy of the 2D $k$-mesh for structural optimization is 1200 $k$-points/{\AA$^{-2}$}. For electron band structure and carrier effective mass, we use a higher-density $k$-mesh (3000 $k$-points/{\AA$^{-2}$}), and an interpolation of the eigenvalues is performed using \textsc{BoltzTraP2}~\cite{Boltztrap,madsen2018boltztrap2}. The interpolated bands are further used to calculate the carrier effective masses.


The phonon dispersions have been calculated using density-functional perturbation theory (DFPT) as implemented in \textsc{Quantum Espresso}~\cite{Giannozzi_2009, Giannozzi_2017}. 
A $\Gamma$ centered k-point grid of  $12\times12\times1$ and a cut-off of 90 Ry were used to converge the ground state charge density. 
We have used a vacuum distance off  15~\AA\ and a 2D Coulomb cutoff (otherwise, a weak longitudinal optical/transverse optical splitting would occur at the $\Gamma$ point for some of the phonon modes).  We computed dynamical matrices on a uniform $4\times4$ $\Gamma$-centered coarse grid and performed a Fourier transform to obtain the interatomic force constants. The interatomic force constants were then used to obtain the phonon dispersions. All calculations with the \textsc{Quantum Espresso} code were performed using the norm-conserving pseudopotentials from the PseudoDojo project~\cite{VANSETTEN201839,SCHLIPF201536}.  Total energy differences and optimized structures are almost identical in calculations with the \textsc{VASP} and \textsc{Quantum Espresso} codes.

For geometry optimization and phonon calculations, we use the Perdew-Burke-Ernzerhof~\cite{pbe} (PBE)
exchange-correlation functional. {We note that by using different functionals for the geometry optimization (in particular the local density approximation, LDA, that tends to slightly overbind), the energetic ordering of simple square and snub-square phases (and also of the additional phases calculated in Ref.~\onlinecite{Gao2021}) may change.}
The electronic band structures are calculated with the He{y}d-Scuseria-Ernzerhof HSE06~\cite{HSE06} hybrid functional. 

To obtain the optical absorption spectra, we start with the energy eigenvalues and Kohn-Sham wave functions obtained via DFT-PBE with the \textsc{Quantum Espresso} code. We
first perform $G_0W_0$ calculations to correct the energy eigenvalues. We use $9\times9\times1$ uniform $\Gamma$ centered $k$-point grids and include 600 Kohn-Sham states to converge the band structure for the materials. A cutoff of 8~Ry is used to construct the dielectric tensor, and a plasmon-pole scheme~\cite{PhysRevLett.62.1169} is employed to model the frequency dependence of dielectric screening. Later, we perform Bethe–Salpeter equation (BSE)~\cite{RevModPhys.74.601} calculations to obtain the absorption spectrum, including electron-hole interactions. We use a $24\times24\times1$ uniform $\Gamma$ centered $k$-point grid to converge the absorption spectra. A total of 400 Kohn-Sham states and a cutoff of 8~Ry (109~eV) is used to build the static dielectric tensor. We include the top four conduction and bottom five valence bands to construct the BSE Hamiltonian and employ the Tamm–Dancoff approximation to decrease the computational cost~\cite{PhysRevB.57.R9385}. In both $G_0W_0$ and BSE calculation, a Coulomb cutoff of 32~Bohr is set along the non-periodic direction to remove the interactions with periodic images~\cite{PhysRevB.73.233103}. Both BSE and $G_0W_0$ calculations are performed using the \textsc{YAMBO} code~\cite{MARINI20091392,Sangalli_2019}.

The calculation of 2D films on a substrate is calculated with a non-local van der Waals corrected functional (optB86b-vdW)~\cite{Klime2011}, and a six-layer slab of Cu-(001) surface is used as substrate. In this case, the three bottom layers of Cu-atoms are held fixed for the geometry optimizations while the remaining atoms can relax.

\section{Results and Discussion}
\label{sec:results}
\begin{table*}[]
    \caption{Summary of the structural and bond order of the calculated 2D-M$_2$Ch structures. We present the in-plane cell parameters ($a$, in both phases $a=b$, in \AA), the distances between the metallic atoms ($D_\text{M--M}$ in \AA), the Ch--M--Ch bond angles ($\theta$ in degree), and the bond orders between adjacent metal atoms (BO$_\text{MM}$). For $P42_{1}2$ structures the relative differences for $D_\text{M--M}$ and $\theta$ to corresponding $P4_nmm$ structures are also shown in parentheses.}
    \label{tab:bond}
    \centering
    \begin{tabular*}{\linewidth}{@{\extracolsep{\fill} } l | c c c c | c c c c }
         & \multicolumn{4}{c|}{$P4/nmm$} & \multicolumn{4}{c}{$P42_{1}2$} \\
         Formula  & BO$_\text{MM}$ & $a$ & $D_\text{M--M}$ & $\theta$  & BO$_\text{MM}$ & $a$ & $D_\text{M--M}$ & $\theta$\\ \hline
\ce{Cu2S}  & 0.02 &  5.179 & 3.63& 180.0 & 0.21 & 5.008 & 2.74 (-24.5\%) & 159.2 (-11.6\%) \\
\ce{Cu2Se} & 0.02 & 5.042 & 3.56 & 180.0 & 0.29 & 4.933 & 2.60 (-27.0\%) & 158.2 (-12.1\%) \\
\ce{Cu2Te} & 0.02 & 5.042& 3.56  & 180.0 & 0.35 & 4.901 & 2.50 (-29.8\%) & 158.2 (-12.1\%) \\
\ce{Ag2S}  & 0.01 & 5.888 & 4.16 & 180.0 & --  & -- & -- & -- \\
\ce{Ag2Se} & 0.01 & 5.904 & 4.17 & 180.0 & 0.10 & 5.784 & 3.43 (-17.7\%) & 165.2 (-8.2\%) \\
\ce{Ag2Te} & 0.01 & 5.947 & 4.20 & 180.0 & 0.27 & 5.719 & 3.06 (-27.1\%) & 159.3 (-11.5\%) \\
\ce{Au2S}  & 0.02 & 5.818 & 4.11 & 180.0 & --   & -- & -- & -- \\
\ce{Au2Se} & 0.02 & 5.788 & 4.09 & 180.0 & 0.34 & 5.579 & 2.95 (-27.9\%) & 157.6 (-12.4\%) \\
\ce{Au2Te} & 0.02 & 5.820 & 4.11 & 180.0 & 0.41 & 5.598 & 2.88 (-30.0\%) & 157.1 (-12.7\%) \\
    \end{tabular*}
\end{table*}

\begin{table*}[]
    \caption{Summary of 2D-M2Ch structures, distance to the convex hull ($E_\text{hull}$ in meV/atom), band gap calculated with the PBE functional (Gap$^\text{PBE}$ in eV) and HSE06 hybrid functional (Gap$^\text{HSE}$ in eV), effective electron mass ($m_\text{e}^*$ in $m_\text{e}$), and light/heavy hole masses ($m_\text{h, L}^*$/$m_\text{h, H}^*$ in $m_e$) at the band edges. For comparison, we showed the space group (Spg.) and PBE band gap (Gap$^\text{PBE}$, taken from the Materials Project database~\cite{Jain2013}) for the experimental 3D crystal structures.}
    \label{tab:summary}
    \centering
    \begin{tabular*}{\linewidth}{@{\extracolsep{\fill} }l| c c c c c c| c c c c c c| l c}
         & \multicolumn{6}{c|}{$P4/nmm$} & \multicolumn{6}{c}{$P42_{1}2$} & \multicolumn{2}{c}{3D Exp.} \\ 
         Formula & 
         $E_\text{hull}$  & Gap$^\text{PBE}$ & Gap$^\text{HSE}$ & $m_\text{e}^*$ & $m_\text{h, L}^*$ &$m_\text{h, H}^*$ &
         $E_\text{hull}$  & Gap$^\text{PBE}$ & Gap$^\text{HSE}$ & $m_\text{e}^*$ & $m_\text{h, L}^*$ &$m_\text{h, H}^*$ & Spg. & Gap$^\text{PBE}$ \\
         \hline\\[-3mm]
         \ce{Cu2S} 
                   &  24 & 0.60 & 1.64 & 0.14 & 0.16 & 1.15 
                   &  13 & 0.16 & 1.07 & 0.12 & 0.14 & 1.11  
                   &  $P4_32_12$ & 0.13
                   \\ 
         \ce{Cu2Se}
                   &  54 & 0.62 & 1.63 & 0.15 & 0.15 & 1.05  
                   &  27 & 0.12 & 1.00 & 0.12 & 0.14 & 0.98  
                   &  $Fm\bar{3}m$ & 0.09
                   \\ 

         \ce{Cu2Te}
                   & 118 & 0.50 &1.38 & 0.14 & 0.14 & 0.64 
                   &  65 & 0.00 & 0.67 & 0.11 & 0.11 & 0.81  
                   & $P6/mmm$ & 0.00
                   \\ 
         \ce{Ag2S} 
                   &  7  & 1.79  & 2.59 & 0.19 & 0.20 & 1.13   
                   &   - & - & - & - & - & - 
                   & $P2_1/n$ & 0.93
                   \\
         \ce{Ag2Se}
                   &   3 & 1.82 & 2.58 & 0.19 & 0.22 & 1.22 
                   &   7 & 1.34 & 2.12 & 0.18 & 0.21 & 1.10  
                   & $P2_12_12_1$ & 0.00
                   \\ 
         \ce{Ag2Te}
                   &  28 & 1.69 & 2.35 & 0.18 & 0.18 & 1.38 
                   &  21 & 0.95 & 1.56 & 0.15 & 0.19 & 1.08  
                   & $P2_1/c$ & 0.00
                   \\ 
         \ce{Au2S} 
                   &  85 & 1.00 & 1.59 & 0.10 & 0.10 & 0.70     
                   & - & - & - & - & - & -  
                   & $P\bar{n}3m$ & 1.91
                   \\
         \ce{Au2Se}
                   &  42 & 1.02 & 1.61 & 0.12 & 0.10 & 0.71 
                   &  57 & 0.00 & 0.42 & 0.07 & 0.07 & 0.74
                   & -- & -- \\ 
         \ce{Au2Te}
                   &  31 & 0.93 & 1.44 & 0.12 & 0.10 & 0.67 
                   &  29 & 0.00 & 0.09 & 0.04 & 0.06 & 0.13 
                   & -- & -- \\
    \end{tabular*}
\end{table*}

\subsection{Structure and Bonding}

As a 3D crystal, \ce{Ag2Se} is naturally found in the form of naumannite, an orthorhombic system with $P2_12_12_1$ space group symmetry~\cite{anthonyeds}. Two inequivalent metallic sites are found in naumannite, namely a 3- and a 4-fold coordination centers~\cite{Jain2013}. \ce{Ag2S} crystalizes in monoclinic anti-\ce{PbCl2}-like structure and transform to \ce{Ag2Se}-like structure at high-pressure\cite{SantamaraPrez2012}. The compound \ce{Ag2Te}~\cite{FRIEH1959} forms in a distorted \ce{ZrSi2}-like structure with the monoclinic space group $P2_1/c$. The structure consists of two inequivalent Ag sites acting as 10-fold and 8-fold coordination centers, respectively. The 3D \ce{Cu2S} material crystallizes in the tetragonal $P4_32_12$ space group~\cite{Janosi1964}, where the copper atoms are coordinated with three sulfur atoms in a trigonal planar configuration. While \ce{Cu2Se} is predicted from theory to form in the same structure as \ce{Cu2S}, experimental reports observe crystalization in cubic phases \cite{Davey1923, Borchert1954}. In contrast, \ce{Cu2Te} can be found uniquely in 2D hexagonal sheets with the space group $P6/mmm$ \cite{Nowotny1946}. It turns out that \ce{Au2S} is the only reported \ce{Au2Ch} compound, exhibiting a cuprite-like structure with the cubic spacegroup $P\bar{n}3m$ \cite{hirsch1966cristallographie}. The S ligands form 4-fold coordination centers with the Au cations in the cuprite-like phase. From this short overview, it is clear that, in the three-dimensional world, metal chalcogenides crystallize in many structures with different coordination and bonding patterns.

The 2D snub-square lattice belongs to the $P42_{1}2$ crystal space group (here, we will use the three-dimensional space group for convenience). It can be seen as the result of a rotation of the tetragonal pyramids of the square lattice  (Fig.~\ref{fig:structure}a) belonging to the space group $P4/nmm$. The rotation causes a distortion of the perfect squared metallic network of the $P4/nmm$ phase. Due to this close relation between the $P42_{1}2$ and the $P4/nmm$ phases, we will often compare them in the following.

In Table~\ref{tab:bond}, we compare the structural parameters among the \ce{M2Ch} systems. For $P42_{1}2$ structures, the degrees of distortion are quantified by the relative difference for Ch--M--Ch bond angles and M--M distances compared to the $P4/nmm$ counterparts. We also computed atom bond orders~\cite{manz2016introducing, manz2017introducing} between the nearest noble metal atoms in these systems. Of note is that for \ce{Ag2S} and \ce{Au2S}, the $P42_{1}2$ structures symmetrize to the $P4/nmm$ lattice during structural relaxation, indicating that for these two systems, the $P42_{1}2$ structure is dynamically unstable.

\begin{figure}[!h]
\begin{tabular}{ccc}
\includegraphics[width=4cm]{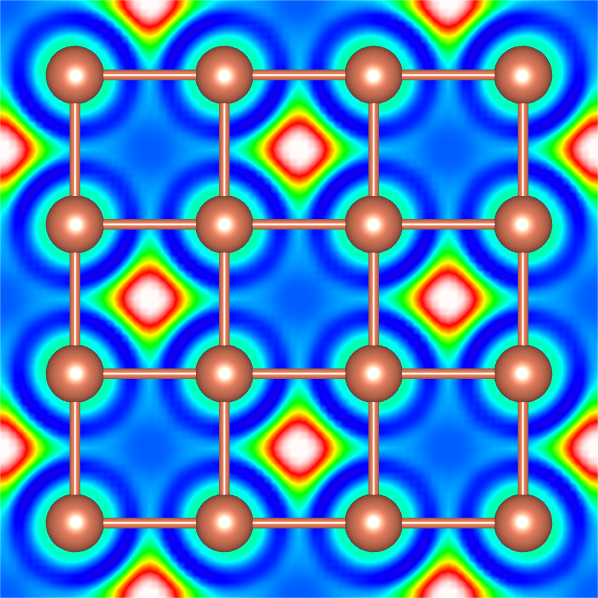} &
\includegraphics[width=4cm]{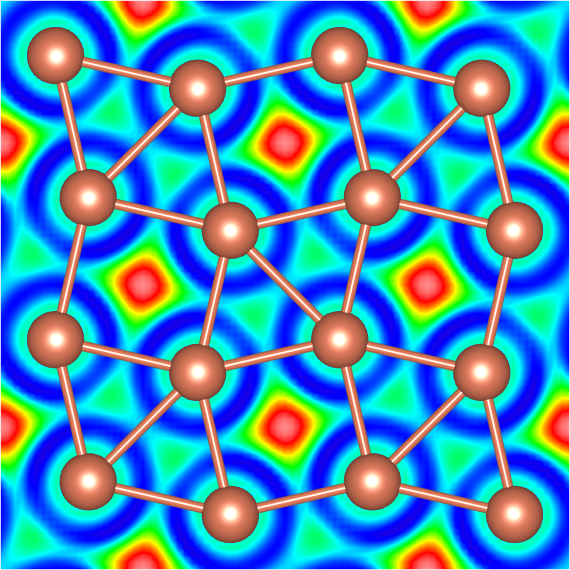}   &
\includegraphics[height=4cm]{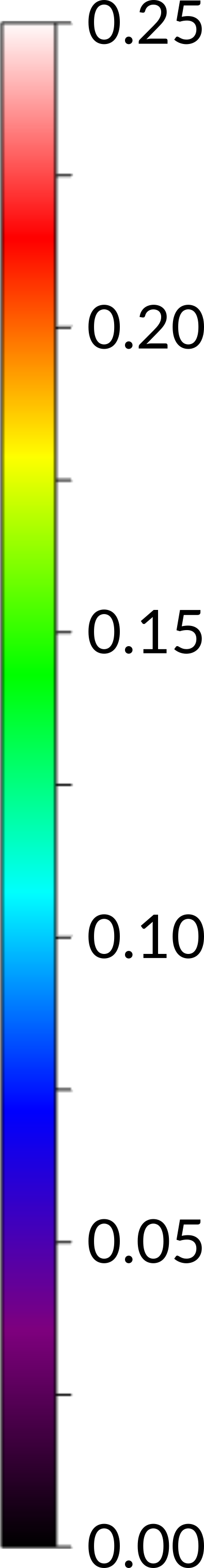}\\
(a) & (b) \\[0.25cm]
\includegraphics[width=4cm]{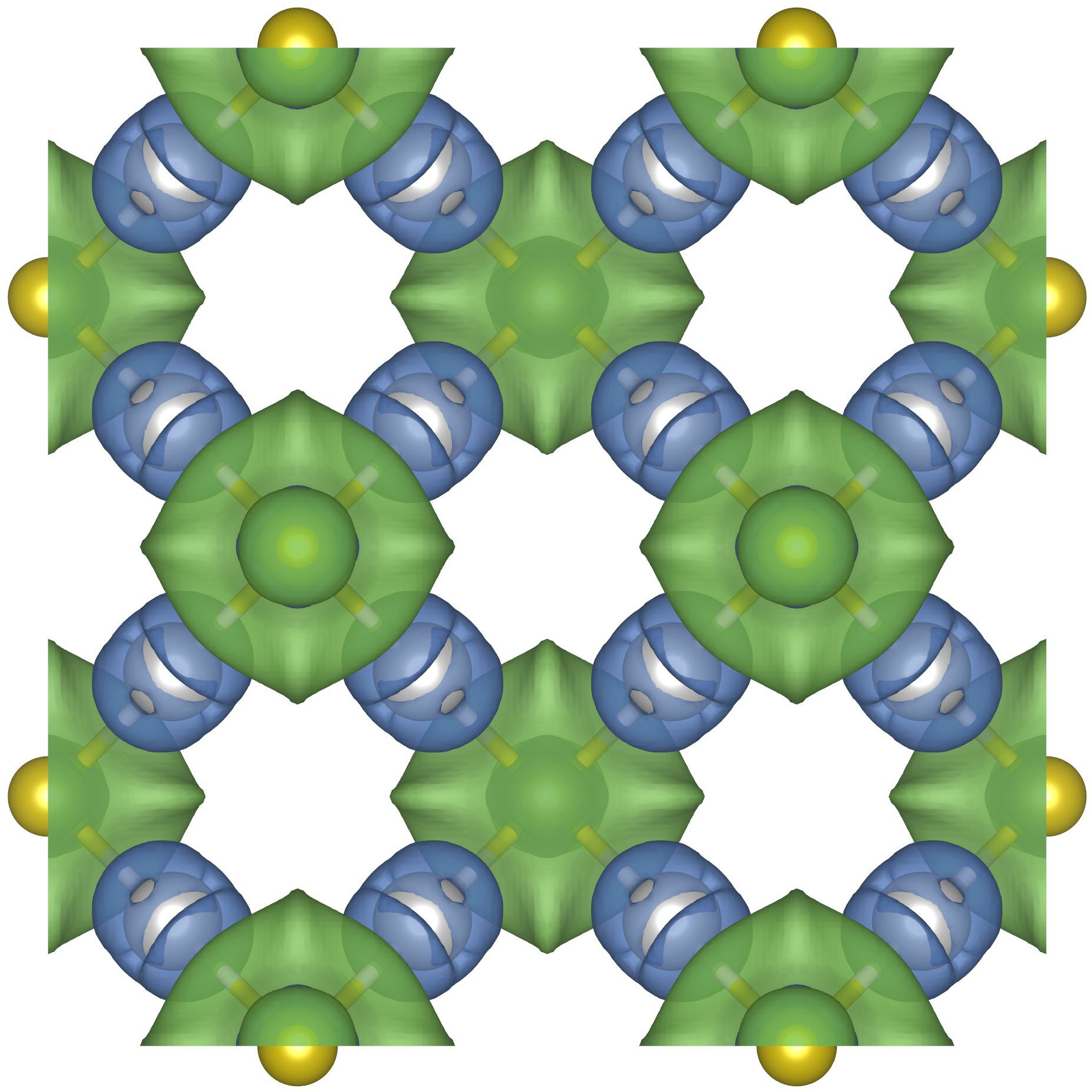} &
\includegraphics[width=4cm]{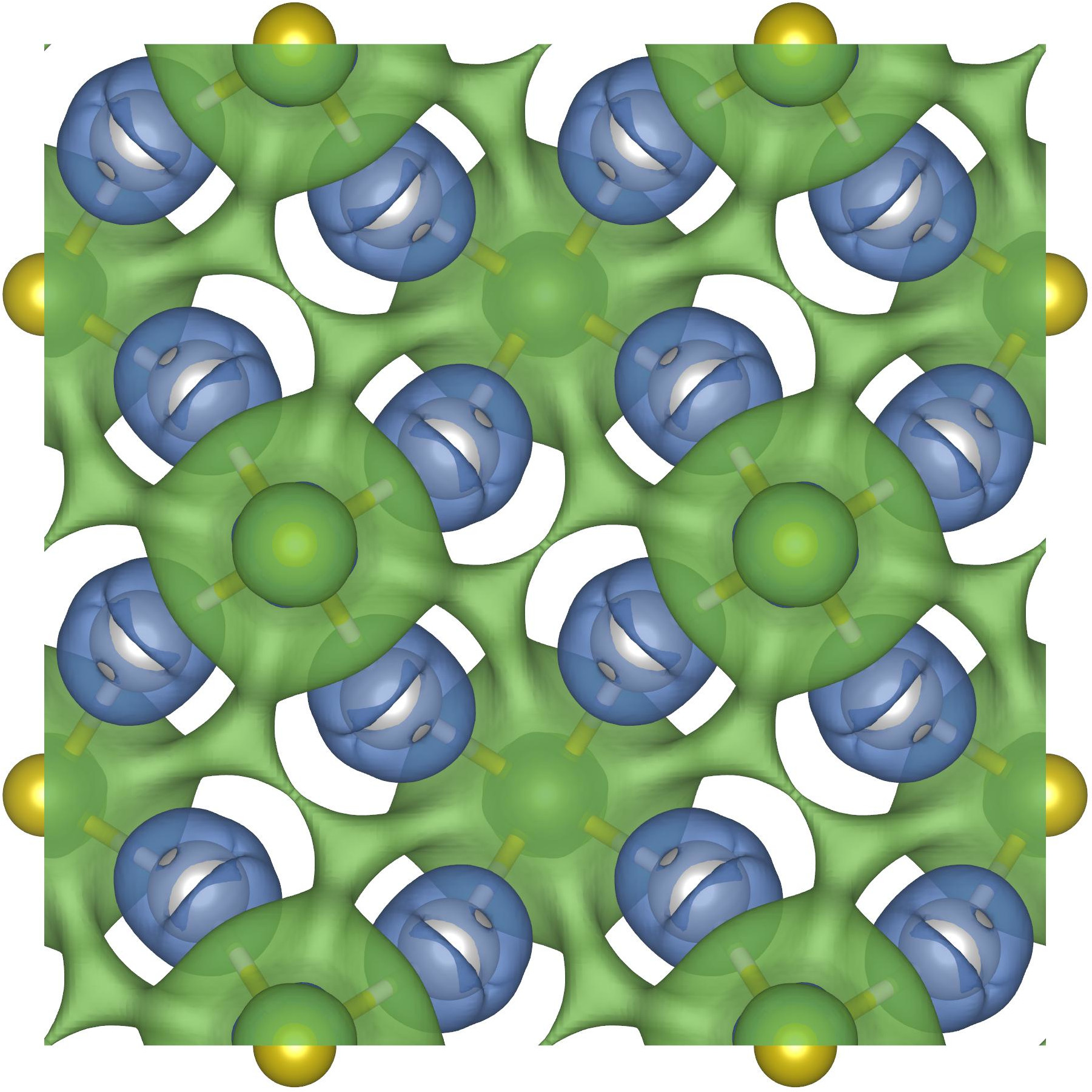} & \\
(c) & (d) \\[0.25cm]
\end{tabular}
\caption{Electron localization function (ELF) of (a) the $P4/nmm$ and (b) the $P42_12$ structures for Cu$_2$S (Cu atoms are denoted as bronze color). Iso-surface plot at a value of $\pm$0.003 electron/\AA$^{-3}$ of charge density difference of (c) the $P4/nmm$ and (d) the $P42_12$ structures for Cu$_2$S, where depletion and accumulation of charges compared to atomic density are represented as naval blue and teal colors, respectively.}
    \label{fig:elf}
\end{figure}

From this table, we can conclude several things. First, in the $P4/nmm$ structures the Ch--M--Ch angles are always 180$^o$ (no distortion), M--M distances are above 3.5~\AA{ }and BO$_\text{MM}$ shows negligible metal-metal bonding interactions, which can also be verified by the electron localization function (ELF) and charge density difference depicted in Fig.~\ref{fig:elf}(a) and (c), respectively. On the contrary, in the $P42_{1}2$ geometry, the M--M bonding is noticeable, as shown by the ELF and charge difference plots in Fig.~\ref{fig:elf}. More importantly, there is a clear correlation between the degree of distortion and the increase of M--M bonding. The distortions are more significant for the gold compound and smaller for silver ones with fixed chalcogen and increase for heavier chalcogens for a given metal. The BO$_\text{MM}$ reaches a maximum of $0.41$ in the case of Au$_2$Te. The ELF in Fig.~\ref{fig:elf}(a) and (b) also shows clearly that there is a strong delocalization of the charge going from the $P4/nmm$ to the $P42_{1}2$ phase.

We also list the thermodynamic stability of the 2D-\ce{M2Ch} in both $P4/nmm$ and $P42_12$ symmetries (see Fig.~\ref{tab:summary}). The system that is furthest from the convex hull is \ce{Cu2Te} at 118~{meV/atom} in the $P4/nmm$ geometry and at 65~{meV/atom} in the $P42_12$ counterpart. The most stable system is \ce{Ag2Se} at a mere 7~meV/atom in $P42_12$ and 3~meV/atom in $P4/nmm$. As mentioned above, only \ce{Cu2Ch} systems stabilize in the $P42_12$ for all three chalcogenides. For both geometries (not applicable for \ce{Ag2S} and \ce{Au2S}), $E_\text{hull}$ increases for copper and silver chalcogenides as the anions get heavier while this trend is reversed among gold compounds.

Interestingly, the relative stability between the $P42_12$ and $P4/nmm$ analogs correlates with the degree of distortion and bond order for copper compounds, showing that in copper analogs the M--M bonding interaction caused by distortion crucially stabilizes the $P42_12$ geometry. However, for silver and gold compounds, the selenides are more stable in the $P4/nmm$ geometry.

\subsection{Electronic properties}
\begin{figure*}[htb]
    \begin{tabular}{ccc}
    \includegraphics[height=5.4cm]{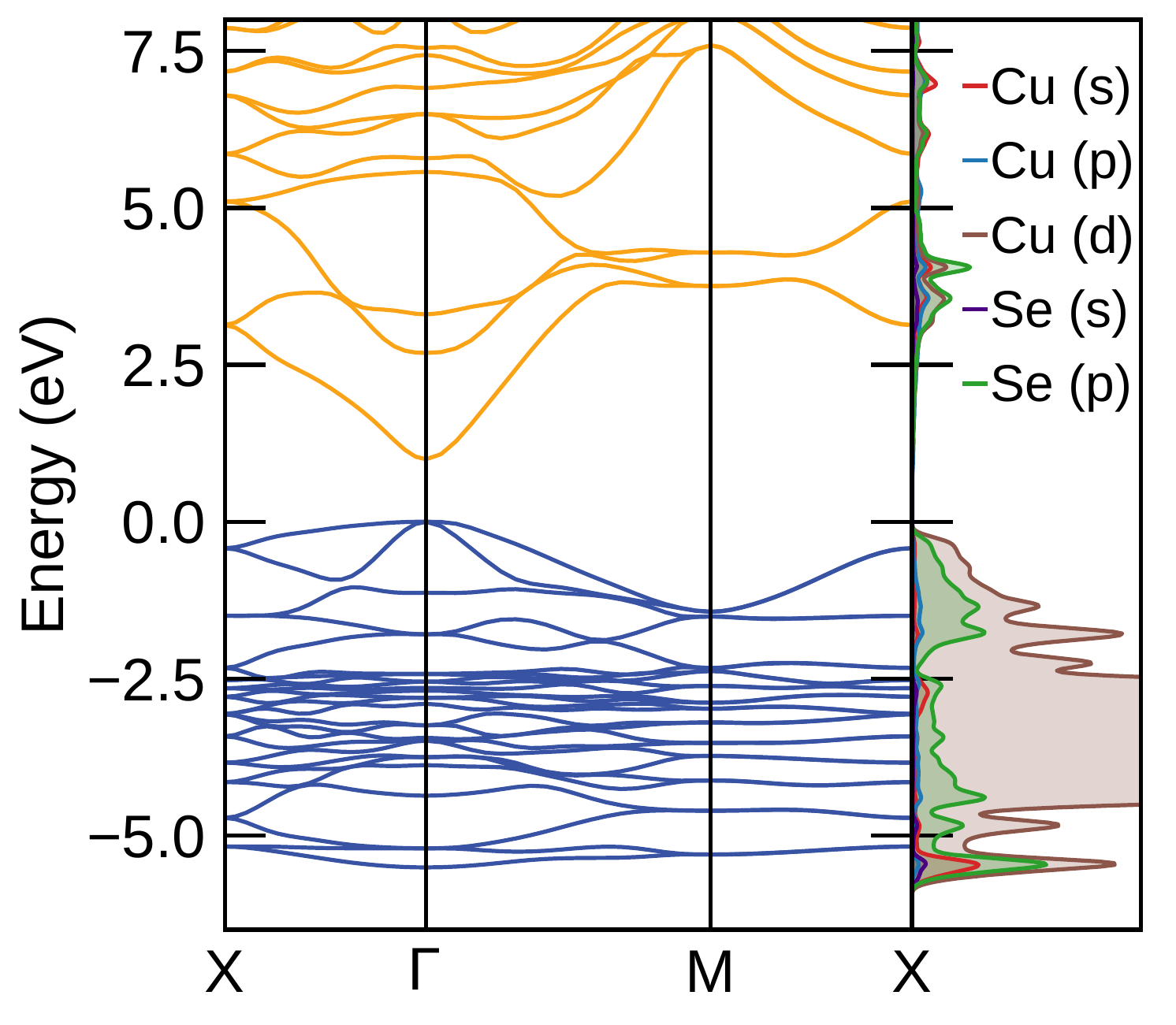}  &
    \includegraphics[height=5.4cm, trim={30 0 0 0},clip]{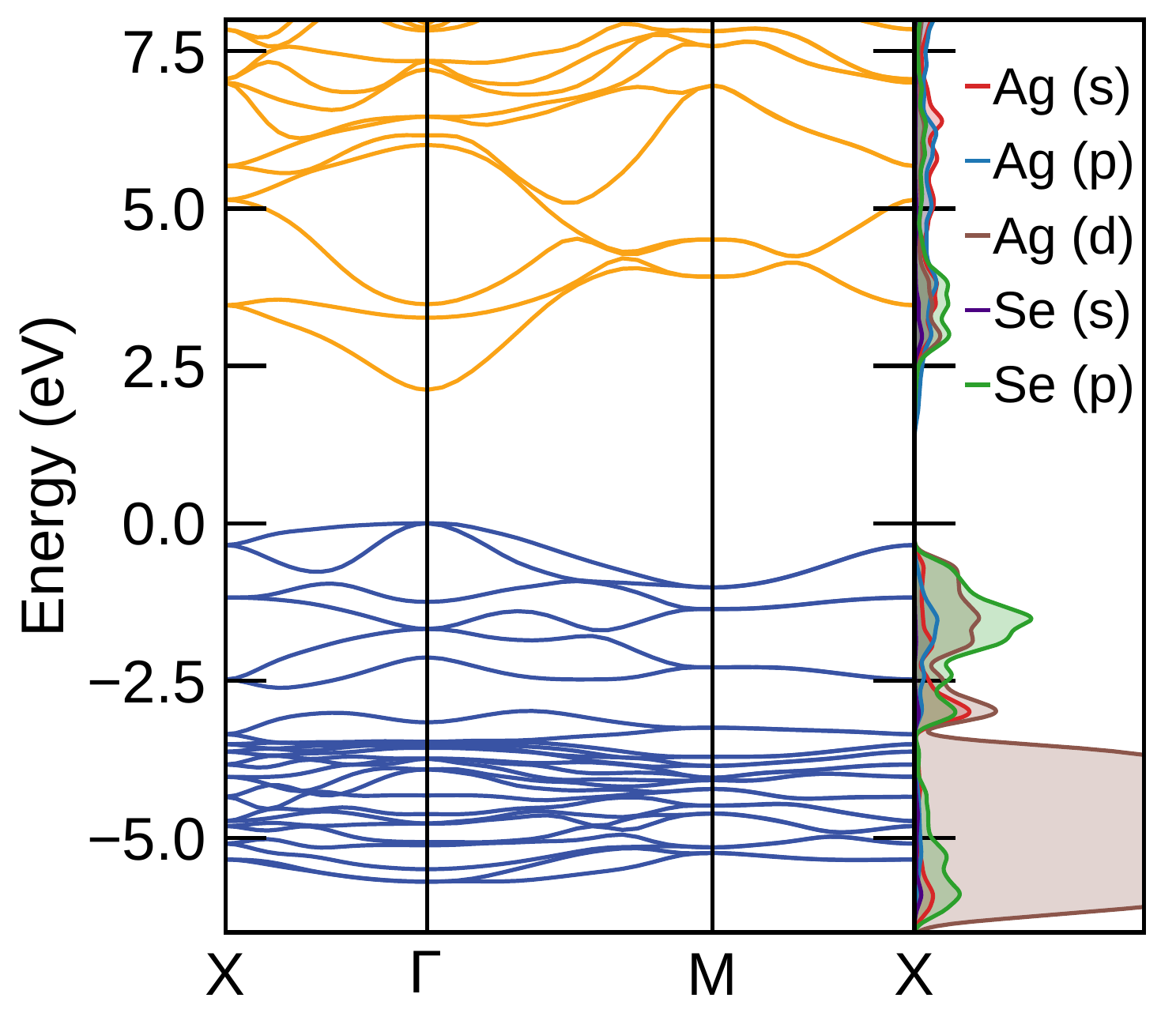} &
    \includegraphics[height=5.4cm, trim={30 0 0 0},clip]{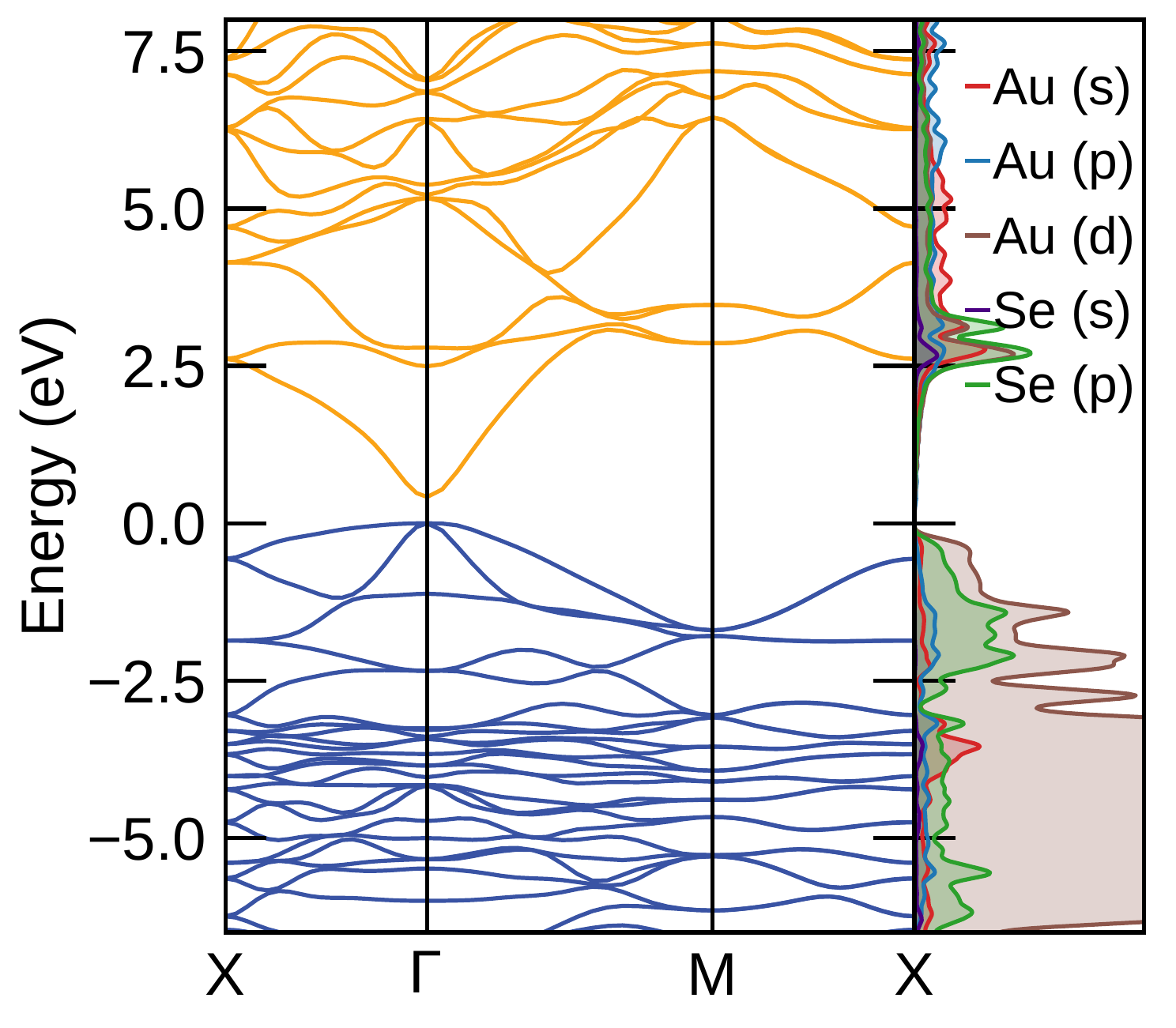} \\[0.25cm]
    (a)~\ce{Cu2Se} & (b)~\ce{Ag2Se} & (c)~\ce{Au2Se} 
    \end{tabular}
    \caption{Electronic band structures of the $P42_12$ phases of (a)~\ce{Cu2Se}, (b)~\ce{Ag2Se}, and (c)~\ce{Au2Se}, calculated with the HSE06 hybrid functional.}
    \label{fig:band}
\end{figure*}

In Fig.~\ref{fig:band}, we show the electronic band structure and the projected density of states for the three selenide $P42_{1}2$ systems computed using the screened hybrid density-functional HSE06~\cite{HSE06}. 

The systems exhibit a direct gap at the $\Gamma$ point. The highest valence states are doubly degenerate with a stark difference in the dispersion behavior near the $\Gamma$ point. 
One of the states is characterized by a nearly flat dispersion curve, while the other shows a very pronounced curvature. Consequently, these systems could accommodate light holes as well as heavy ones. The lowest conducting states show somewhat curved dispersion curves around $\Gamma$ compatible with particles of effective masses similar to those of the aforementioned light holes. Table~\ref{tab:summary} lists the band gap and the particle/hole effective masses for the $P42_12$ systems and their $P4/nmm$ analogs.

It is well known that the PBE functional underestimates considerably band gaps~\cite{Tran2021}. In fact, three of the $P42_12$ systems were misidentified as metals at the PBE level. The electronic band gaps at the HSE06 level show that $P4/nmm$ systems are moderate- to wide-band-gap semiconductors with band gap values from 1.59 up to 2.59~eV. All the $P42_12$ systems have smaller band gaps than their previously mentioned counterparts, with values ranging from 90~meV to 2.12~eV, consistent with the larger degree of delocalization of the electronic states in these systems.

As expected, moving down the chalcogen group for a specific metal reduces the gap in both phases. We also note that the heavier the chalcogen, the more significant the HSE06 band gap difference ($\Delta$Gap) between $P4/nmm$ and $P42_12$ phases, with $\Delta$Gap largest in the case of \ce{Au2Te}. Clearly, there is also a strong positive correlation between $\Delta$Gap and the distortion/M--M bonding order. The correlation is consistent with the ELF and charge transfer shown above, as forming the M--M bond effectively reduces the charge transfer from metal to chalcogen, weakening the covalent M--Ch bonding and consequently reducing the gap.

We can see that light and heavy holes appear in the studied systems in both geometric configurations. The transformation from one space group to the other has limited impact on the particle/hole effective masses, except in the case of \ce{Au2Te}, where we find the heavy hole effective mass to be reduced to less than a fifth of its original value. The band masses of these systems show considerable improvement in the $m_e^*$ over commercialized n-type TCOs~\cite{Hautier2014} to below 0.2 $m_0$ across all compositions and in both polymorphs. The light holes show similar improvement, reducing $m_h^*$ to below 0.2~$m_0$. However, the heavy holes remain a potentially limiting factor for functional p-type mobility. Unfortunately, the maximum band gap we find in our chalcogenides (2.59~eV for \ce{Ag2S}) is well within the visible spectrum, limiting the usability of these materials as $n$- or $p$- type transparent conductors for transparent electronic applications\cite{minami_2000,afre2018transparent,Hautier2013effectivemass}.

\subsection{Optical absorption}

In this section, we examine the excitonic effects in the optical absorption of $P4/nmm$ and $P42_12$ phases of \ce{Ag2Se}. The optical response of a typical two-dimensional semiconductor is dominated by excitons due to reduced environment screening~\cite{RevModPhys.90.021001}. To describe the optical absorption spectrum, we calculate the imaginary part of the dielectric tensor, which is given by~\cite{PhysRevB.62.4927}
 \begin{equation}
     \varepsilon_2(\omega) = \frac{8 \pi^2 e^2}{\omega^2}\sum_{S} \Big\lvert \sum_{kcv} A_{kcv}^S\mathbf{e}\cdot \langle vk |\mathbf{v}|ck\rangle \Big\lvert ^2 \delta(\omega-E_S)
 \end{equation}
where $\mathbf{e}$ is the light polarization direction, $\mathbf{v}$ is the velocity operator, $A_{kcv}^S$ are the expansion coefficients of the exciton eigenstates, calculated in the electron-hole basis with the help of the Bethe-Salpeter-Equation, and $E_S$ are exciton energies.

\begin{figure}[!h]
\includegraphics[width=\columnwidth]{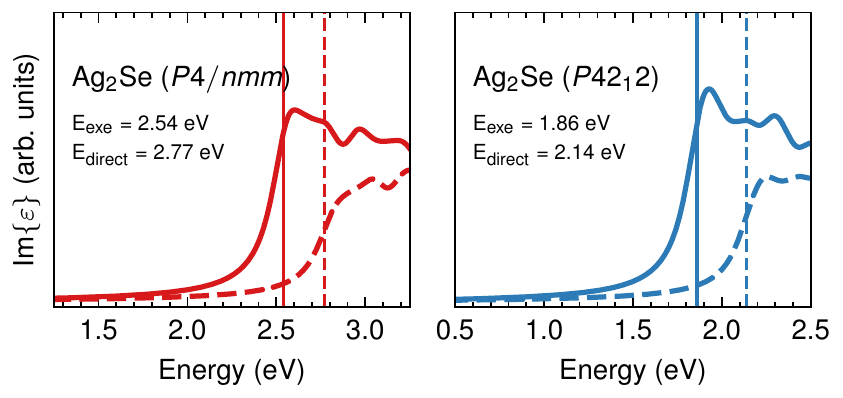} 
\caption{\raggedright Optical absorption spectra of (left) $P4/nmm$ \ce{Ag2Se} and (right) $P42_12$ \ce{Ag2Se} with (solid lines) and without (dashed lines) electron-hole interaction. Vertical solid and dashed lines denote the first exciton energy ($\text{E}_{\text{exe}}$) and the direct band gap  ($\text{E}_{\text{direct}}$) respectively.}
    \label{fig:bse}
\end{figure}

In Fig.~\ref{fig:bse}, we show the absorption spectra computed along the in-plane direction for the $P4/nmm$ and $P42_12$ phases of \ce{Ag2Se}. The vertical solid and dashed lines denote the first exciton energy ($\text{E}_{\text{exe}}$) and the direct band gap  ($\text{E}_{\text{direct}}$) calculated with $G_0W_0$, respectively. The absorption on-sets shift by $\sim$0.3~eV when the electron-hole interaction is included, indicating relatively strong excitonic effects in both phases of \ce{Ag2Se}. The first exciton in both phases is optically bright. It is doubly degenerate with different effective masses (meaning that in the exciton dispersion, for finite wave vector $\mathbf{k}$, there are two different branches). It mainly consists of transitions from the top valence bands to the lowest conduction band at the Brillouin zone center.

\begin{figure}[h]
\includegraphics[width=0.95\columnwidth]{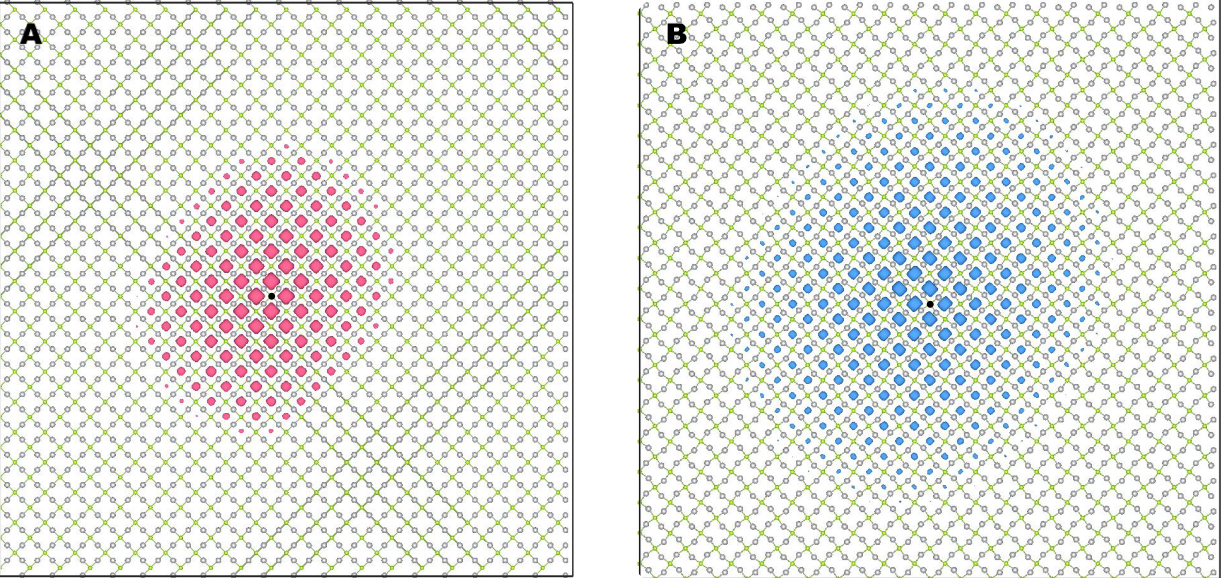} 
\caption{The total probability density of the first bright exciton of (a)~$P4/nmm$ and (b)~$P42_12$ phases of \ce{Ag2Se}, the hole is fixed on the Se atom denoted by a black circle. Green and grey circles denote Se and Ag atoms, respectively.}
\label{fig:bse_exe_wf2}
\end{figure}

In Fig.~\ref{fig:bse_exe_wf2}, we plot the total probability density of the first exciton for both phases of \ce{Ag2Se}. We observe that both excitonic wave functions are rather extended in real space, showing that the exciton is of the Wannier-Mott type. The extension is considerably higher for the $P42_12$ phase, which is compatible with the lower band gap, the increased screening, and, consequently, with the lower excitonic binding energy. 

\subsection{Phonons}

\begin{figure}[!h]
\includegraphics[width=0.95\columnwidth]{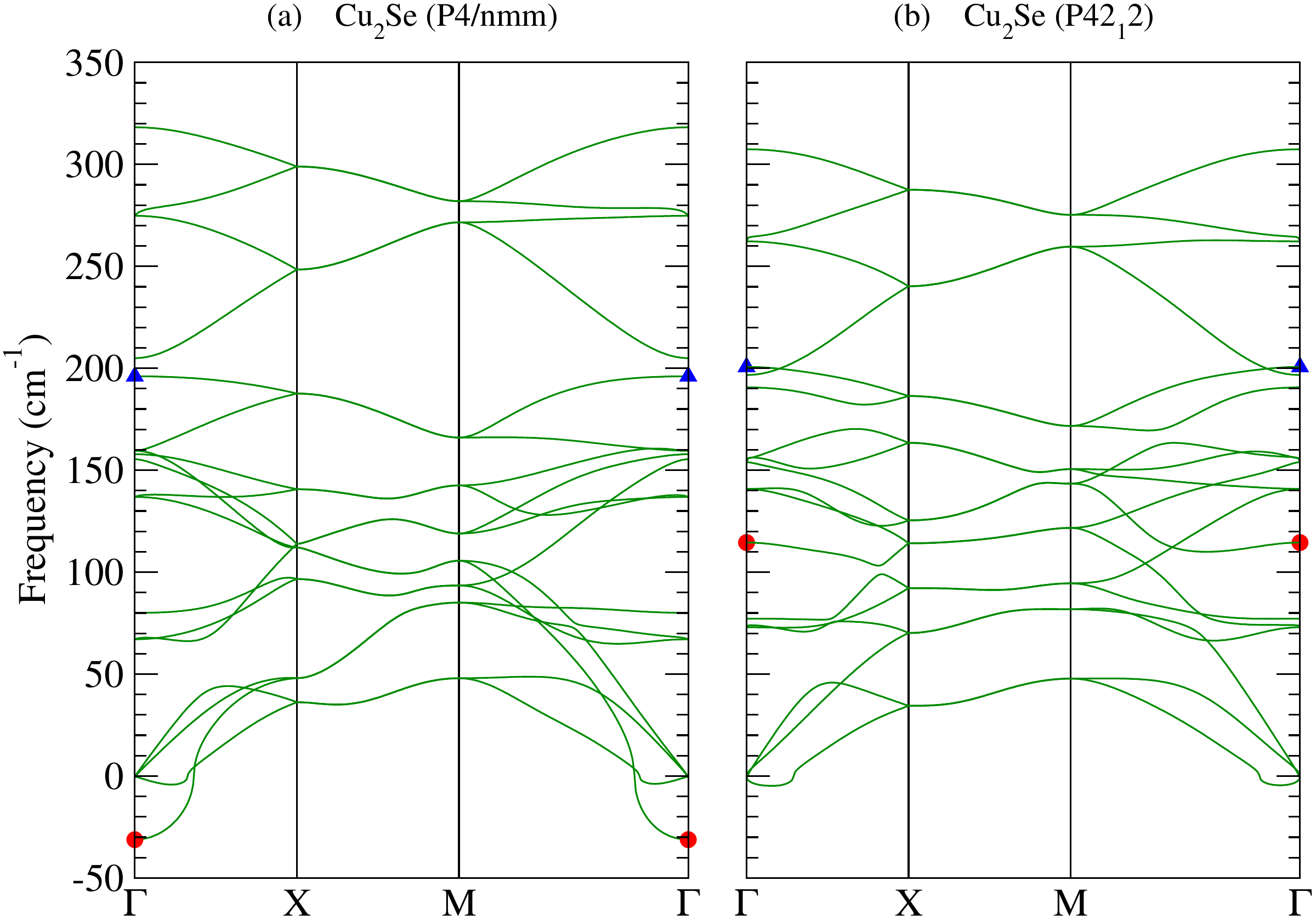} 
\caption{Calculated phonon dispersion of \ce{Cu2Se} (a) in the $P4/nmm$ phase and (b) in the $P42_12$ phase. The soft mode of the $P4/nmm$ phase at $\Gamma$ (marked by red circle) is responsible for the transition to the $P42_12$ phase and acquires there a finite frequency of 114.5 cm$^{-1}$. The blue triangles mark the Raman active A$_{1g}$ (A$_1$) mode in the $P4/nmm$ ($P42_12$) phase, respectively. (The mode eigenvectors are displayed in panels (b) and (c) of Fig.~\ref{fig:raman}).}
\label{fig:phonon}
\end{figure}

\begin{figure}[!h]
\includegraphics[width=0.95\columnwidth]{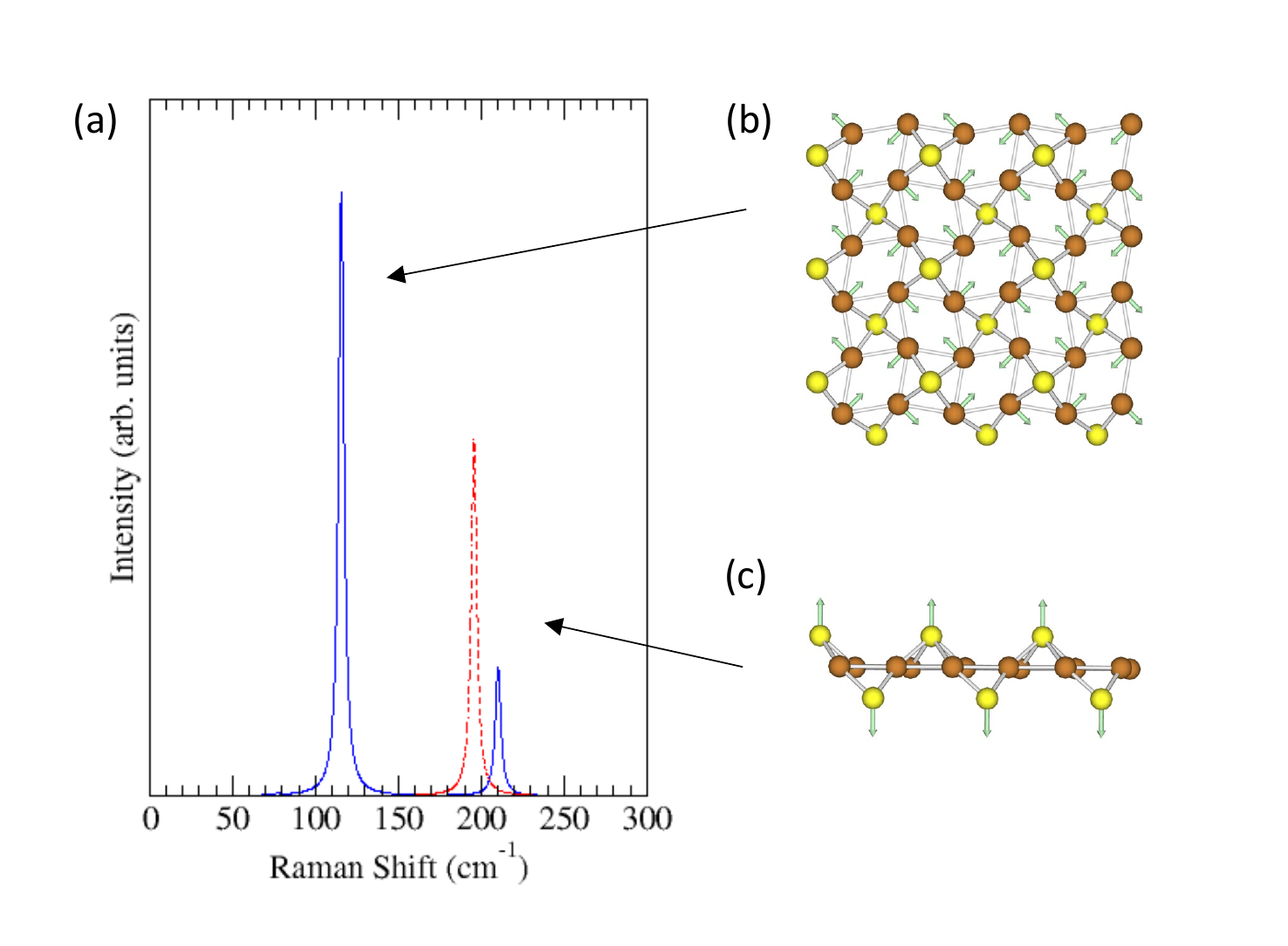} 
\caption{(a) Calculated Raman spectrum of $P42_12$ \ce{Cu2Se} (blue solid line) and of $P4/nmm$ \ce{Cu2Se} (red dashed line). (b) Sketch of the vibrational mode responsible for the Raman peak at 120 cm$\-1$ (and representing the soft mode in the $P4/nmm$ phase). (c) Sketch of the A$_{1g}$ mode.}
\label{fig:raman}
\end{figure}

In Fig.~\ref{fig:phonon} we present the phonon dispersion for \ce{Cu2Se} in the $P4/nmm$ and in the $P42_12$ phases. The main difference between the two dispersions is the soft mode at $\Gamma$ for the $P4/nmm$ phase. The phonon eigenvector corresponding to this soft mode is displayed in panel (b) of Fig.~\ref{fig:raman}. This ``snub-square rotation mode'' drives the symmetry reduction from the $P4/nmm$ to the $P42_12$ phase. The formation of Cu-Cu bonds in the snub-square geometry (as demonstrated by the bond-order calculations in Table~\ref{tab:bond}) is the reason why this mode has imaginary frequency and thus describes the relaxation to the lower-symmetry phase. In the $P42_12$ phase, the same mode exists, but it has a finite (positive) frequency, describing the snub-square rotation around the new equilibrium position.

{We note that the out-of-plane acoustic mode displays a small negative overshoot around $\Gamma$ (for both phases). This is not a real instability but related to numerical inaccuracies in the determination of the equilibrium lattice constant and phonon calculations. Stretching the lattice constant would render this branch entirely positive and give it a linear slope around $\Gamma$. Squeezing the lattice constant increases the negative (imaginary) overshoot and corresponds to long-wavelength wrinkles of the 2D layer.}

\begin{table}[]
    \centering
    \caption{Irreducible representation labels, Infrared (I) or Raman (R) activity, and frequencies (in cm$^{-1}$) of optical phonon modes for \ce{Cu2Se} ($P4/nmm$) and \ce{Cu2Se} ($P42_{1}2$) at $\Gamma$ point. Note that the E modes are doubly degenerate at $\Gamma$.}
    
    \label{tab:irrep}
    \begin{tabular}{cccccccccccc}
    \multicolumn{12}{c}{\ce{$P4/nmm$}} \\
     Label    & A$_{1u}$ & E$_{u}$ & B$_{1u}$ & E$_{u}$ & A$_{2u}$ & B$_{2u}$ & E$_{g}$ & A$_{1g}$ & B$_{1u}$ & E$_{u}$  & A$_{2u}$  \\
    \hline \\[-3mm]
     Activity                      &    

-        & I       &  -        & I      &  I       & -        & R        & R        & -     & I        & I      \\
    Frequency                     & -31       & 67      & 80        & 137    & 155      & 158      & 160      & 196      & 205   & 275      & 318     \\ \\
    \multicolumn{12}{c}{\ce{$P42_{1}2$}} \\
     Label      & E & B$_1$ & A$_1$ & E  & E      & A$_2$ & B$_1$ & B$_2$ & A$_1$ & E   & A$_2$ \\
    \hline \\[-3mm]
     Activity                      & I/R & R   & R     & I/R & I/R  & I     & R     & R     & R     & I/R & I      \\
     Frequency                     & 73  & 77  & 115   & 141 & 154  & 156   & 191   & 197   & 201   & 262   & 307     \\

    \end{tabular}
\end{table}

In Table~\ref{tab:irrep}, we list all modes of the two phases of \ce{Cu2Se} along with their infrared (IR) 
or Raman (R) activity according to group theory.
In Fig.~\ref{fig:raman}, we show the calculated non-resonant Raman spectra\footnote{The Raman spectrum has been calculated with \textsc{Quantum Espresso} using the implementation of Lazzeri and Mauri for the Raman tensor\cite{Lazzeri2003Jan}. Since this implementation is currently only available in the LDA approximation for the exchange-correlation functional, we used LDA for the Raman tensor, but used the phonon frequencies and eigenvectors as calculated with the PBE functional in order to be coherent with the other calculations in this manuscript.}$^,$~\cite{Lazzeri2003Jan} of the two phases of \ce{Cu2Se}. In the spectrum of the undeformed square lattice (red-dashed line), the $A_{1g}$ mode at 196 cm$^{-1}$ dominates the spectrum. The mode consists of vertical (out-of-plane) vibrations of the sulfur atoms (panel (c)) while the Cu atoms are not moving. This mode is similar to the Raman active A$_1$ mode in monolayer MoS$_2$\cite{PhysRevB.84.155413} where the sulfur atoms are also vibrating {in the direction normal to the plane} while the Mo atoms are not moving. Contrary to MoS$_2$, however, the spectrum of $P4/nmm$ \ce{Cu2Se} is a quasi-one peak spectrum where the doubly degenerate $E_g$ mode at 160 cm$^{-1}$ has vanishing intensity and is not visible in the spectrum. 

The spectrum changes to a quasi-two peak spectrum in the $P42_12$ phase (blue line): The ``snub-square mode'' (panel (b)) which is responsible for the instability of the $P4/nmm$ phase acquires a finite frequency of 114.5 cm$^{-1}$. It becomes Raman active and dominates the spectrum besides the high-frequency $A_1$ mode that slightly up-shifts in position. The other Raman active modes, listed in Table~\ref{tab:irrep} have comparatively low intensity.  The Raman spectrum thus gives a clear and easy way to distinguish between the two phases of \ce{Cu2Se}. For iso-structural M$_2$X monolayers that are stable in the $P4/nmm$ phase, the snub-square rotation mode has finite frequency, but is not Raman active due to its A$_{1u}$ symmetry. We thus conclude that for the other elemental combinations discussed in this manuscript, the same two-peak structure serves as a clear signal for the presence of the snub-square deformation. 


\subsection{Substrates}

\begin{figure}[htb]
    \begin{center}
    \begin{tabular}{c}
    \includegraphics[width=6cm]{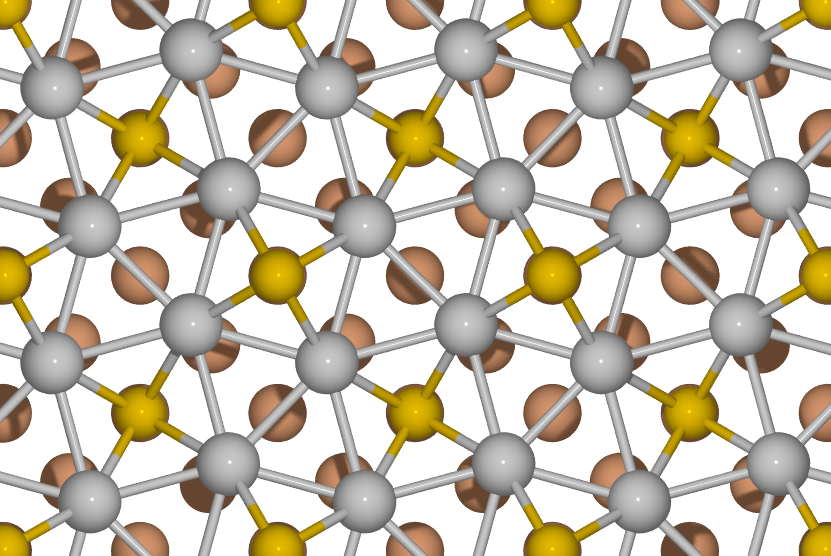} \\[0.5cm]
    \includegraphics[width=6cm]{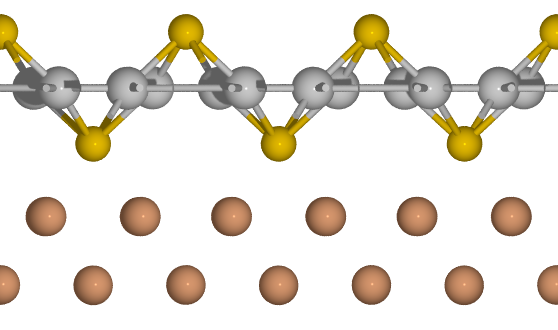} \\
    \end{tabular}
    \end{center}
    \caption{Structures of 2D snub-square \ce{Cu2Se} on Cu (001) substrate. The silver, yellow, and brown spheres denote Cu in \ce{Cu2Se}, Se, and Cu of substrate atoms, respectively. In side view only two out of the six layers of the substrate are shown.}
    \label{fig:surface_structure}
\end{figure}

To investigate possible substrates suitable for synthesizing the snub-square lattice, { we first searched through all simple elementary crystals and binary oxides for potential substrates, and we found 76 substrates with a lattice mismatch below 5\%. We then looked at these, searching for substrates that matched the symmetry of the 2D layer. This led us to choose Cu (001), Ge (001), Pt (001) for \ce{Ag2S}, and Cu (001), Ge (100), Pd (001) for \ce{Cu2Se}. After geometry optimization only the Cu (001) substrate preserved, to a large extent, the symmetry of the snub-square lattice}, as shown in Fig.~\ref{fig:surface_structure}. { 
 In the other cases, the 2D layer deformed significantly due to the strong interaction with the substrate.

Furthermore, for \ce{Cu2Se} on Cu(001) substrate,} the average Cu--Se bond length in the 2D film is stretched by 1.8\% to 2.40~\AA, and the average Cu--Se--Cu bond angle is changed slightly to 157.6$^{\circ}$. The bottom Se layer is separated 2.25~\AA\ from the substrate, and the distance from Se to the substrate Cu atoms is 2.89~\AA, much longer than Cu--Se bond length in the 2D film, indicating very weak bonding between the film and the substrate.

{We then calculated the adhesion energies for \ce{Cu2S}, \ce{Ag2S}, \ce{Au2S}, and \ce{Cu2Se} on Cu (001). The results are 55, 71, 90, and 19~meV/\AA$^2$, respectively. }
{The adhesion energy for \ce{Cu2Se} is} within the range of physical adhesion, and is comparable to the adhesion energy of graphite (about 26 meV/\AA$^2$)\cite{HAN201948}. Therefore, it might be possible to obtain a free-standing \ce{Cu2Se} layer via mechanical exfoliation of deposited layers on the Cu-substrate.
{However, for the sulfides, Cu(001) exhibits a stronger bond with the films and is less ideal for applying mechanical exfoliation to obtain free-standing layers.}


\begin{figure}[htb]
  \includegraphics[width=9cm]{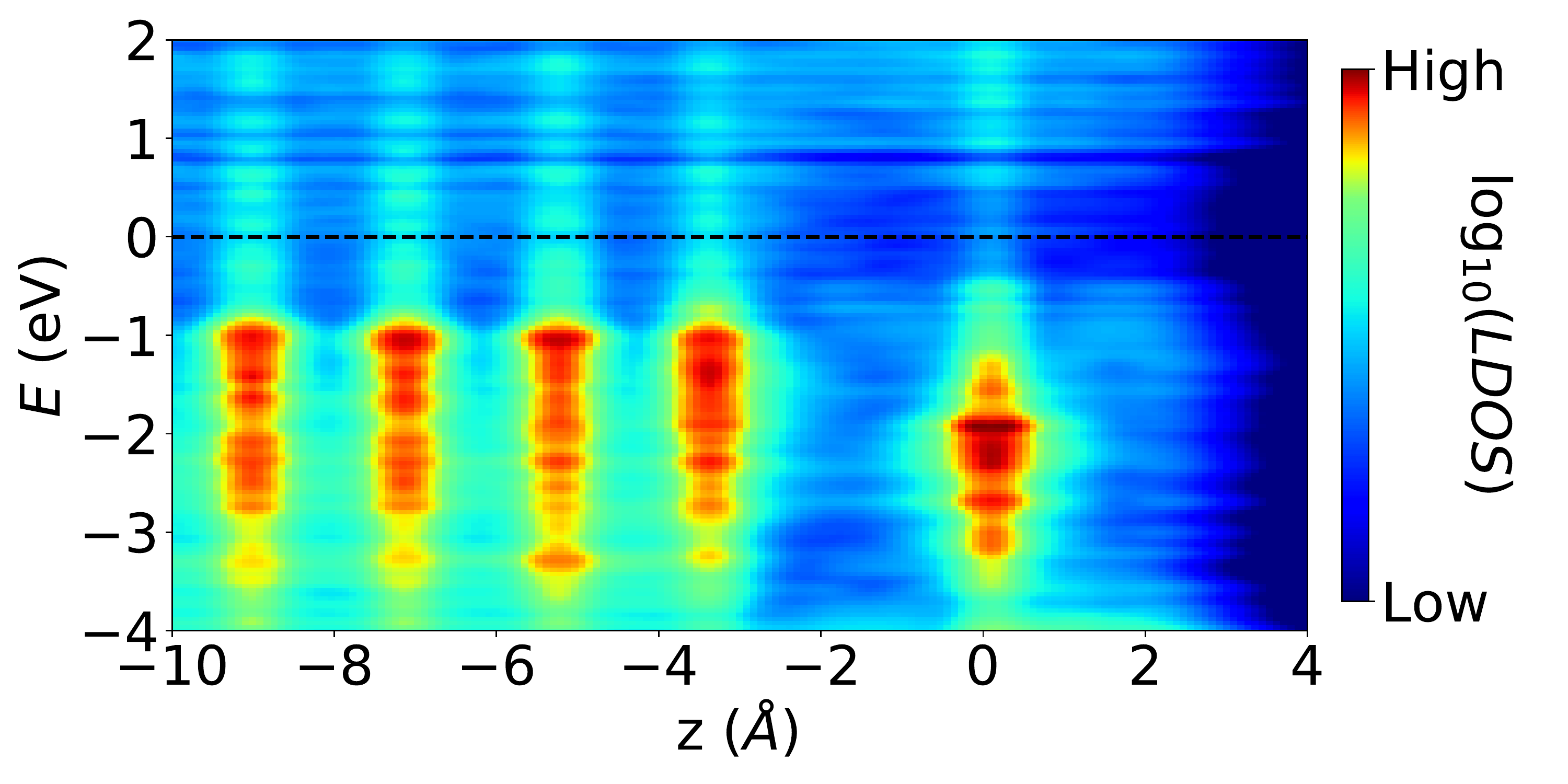}
  \caption{The average local density of states (LDOS) for each $(00z)$ plane for 2D snub-square \ce{Cu2Se} on Cu (001) substrate. the Fermi level is shifted to 0~eV, and the 2D-\ce{Cu2Se} layer is located at $z = 0$~\AA}
  \label{fig:surface_dos}
\end{figure}

In Fig.~\ref{fig:surface_dos} we explore further the interaction between film and substrate by plotting the plane averaged density of states as a function of the plane distance using \textsc{DensityTool}~\cite{Lodeiro2022}. Clearly, a small part of 3d-states from the Cu-substrate is located in the middle of the gap of the film and there is only small mixing between the states of the substrate and film, consistent with the small adhesion energy.

\subsection{Quasicrystals}
Two-dimensional quasicrystals were discovered experimentally in \ce{BaTiO3} on top of Pt and a few other related systems~\cite{Frster2013,Zollner2019}. Changing the synthesis conditions, it was also possible to create simpler approximant structures, periodic 2D crystals that can be inflated by a recursive approach to generate the quasi-crystalline system. The stability of the \ce{M2Ch} snub-square lattice, which can be seen as a small approximant structure, raises, therefore, the interesting question if noble metal chalcogenides quasicrystals are possible. 

It is straightforward to generate larger and larger approximants for our system. However, we are immediately faced with two difficulties: (i)~the ratio of squares and triangles changes during the inflation process, and tends to an irrational number in the quasicrystalline limit. This poses the problem of charge neutrality, as the balance of the positive metal charges is no longer compensated by an equal number of negative chalcogenide charges. A possible solution is electron transfer from a metallic substrate to make up for the unbalanced charge, or by forming defects (e.g., vacancies) in the 2D structure. (ii)~The chalcogenide atoms in our snub-square structure are out-of-plane and show an alternation that can be seen from the lower panel of Fig.~\ref{fig:surface_structure}. Unfortunately, the inflation structure disrupts this alternation resulting in a frustrated system with two neighbor chalcogen atoms placed either above or below the plane. We can expect this to raise the energy of the system by an amount that clearly depends on the specific chemistry. This situation can also be alleviated by creating chalcogen vacancies in the structure.

To test this hypothesis, we performed DFT calculation in the first inflation of the snub-square structure (of composition \ce{Ag15Se6}). As expected, the steric hindrance of the neighboring Se leads to structural instability that completely destroys the snub-square lattice. We tried to remedy this by removing the Se atom from the central square, but the structure was again highly unstable. As such, it seems very unlikely that a quasicrystal can ever be achieved in this system.

\section{Conclusion}

In this paper, we discussed the snub-square tiling, and its parent square lattice, for a series of noble-metal chalcogenides. We showed that snub-square tiling is closely related to regular square tiling, with a rotation of squares forming the extra metal-metal bond in the former. The metal-metal bonding, leading to a substantial delocalization of the charge, is the key to understanding the structural distortion and the thermodynamic stabilization of the snub-square concerning its square counterpart. It is also responsible for reducing the band gap of the snub-square systems. The valence band edge at $\Gamma$ is doubly degenerate, and the curvature {is different for} these two bands, leading to heavy and light holes. The holes have comparable effective mass to \ce{CuI}, the most promising $p$-type transparent conductor. Combined with the relatively large band gap of some of the chalcogenide systems, the low electron effective mass could be helpful to develop $n$- and $p-$type transparent semiconductors. 

Due to the 2D geometry, the excitonic interaction plays a crucial role in the optical absorption spectra, as expected. The first exciton is bright, and the absorption on-set is largely red-shifted by around 0.3~eV due to the strong exciton binding. The exciton is highly localized at $\Gamma$.

{The square geometry and the snub-square geometry are related by a phonon mode in which the squares formed by the metal atoms get tilted. This mode is soft for the materials where the snub-square geometry is lower in energy than the square one. Upon the snub-square deformation, it acquires finite frequency and its prominent Raman peak is a clear fingerprint of the snub-square geometry.}

Finally, we explored possible substrates that could be used for the experimental synthesis of the snub-square lattice. We find that Cu (001), with a 3\% mismatch with the \ce{Cu2Se} 2D layer, is a good candidate, with a low adhesion energy for mechanical exfoliation. We also tried to construct quasicrystals derived from the snub-square tiling through inflation. However, quasicrystal approximants turned out to be highly unstable due to deviations from charge neutrality and steric hindrance due to frustration.

All these results suggest that noble metal chalcogenide snub-square lattices are very good candidates for experimental synthesis, being very close to thermodynamical stability and compatible with simple surfaces of common metals. Moreover, they exhibit interesting properties and can open a new playing ground for studying frustration in two-dimensional systems.

\section{Data availability}
The relevant data are available at Materials Cloud (\url{https://doi.org/10.24435/materialscloud:sb-cy}). The structures, distances to the hull, and other basic properties, can be accessed at \url{https://tddft.org/bmg/physics/2D/} through a simple web-based interface.

\section{Supporting Information}
Electronic band structures and phonon band structures for all studied systems.

\section{Acknowledgements}

This research was funded in part, by the Luxembourg National Research Fund (FNR), Inter Mobility 2DOPMA, grant reference  15627293. We also acknowledge the computational resources awarded by XSEDE, a project supported by National Science Foundation grant number ACI-1053575. The authors also acknowledge the support from the Texas Advances Computer Center (with the Stampede2 and Bridges supercomputers). We also acknowledge the Super Computing System (Thorny Flat) at WVU, which is funded in part by the  National Science Foundation (NSF) Major Research Instrumentation Program (MRI) Award \#1726534, and West Virginia University. AHR also recognizes the support of West Virginia Higher Education Policy Commission under the call Research Challenge Grant (RCG) program. MALM gratefully acknowledges the computing time provided to them on the high-performance computers Noctua 2 at the NHR Center PC2. These are funded by the Federal Ministry of Education and Research and the state governments participating on the basis of the resolutions of the GWK for the national highperformance computing at universities (www.nhr-verein.de/unsere-partner). For the purpose of open access, the authors have applied a Creative Commons Attribution 4.0 International (CC BY 4.0) license to any Author Accepted Manuscript version arising from this submission. 

\section{Competing Interests}
The authors declare no competing financial or non-financial interests.

\section{Author  Contributions}
AHR generated the crystal structures; MALM, HCW and AWH performed the DFT calculations for energies, band structures, and interaction with substrates; MN and LW performed optical property and phonon calculations; AHR, LW, and MALM directed the research; all authors contributed to the analysis of the results and to the writing of the manuscript.

\bibliography{bib}

\end{document}